\pdfoutput=1 
\documentclass[technical]{adada}
\usepackage{graphicx} 
\usepackage{color} 
\usepackage{url}
\usepackage{xurl}
\usepackage[T1]{fontenc}
\usepackage{amsmath}
\usepackage{array}
\usepackage{booktabs}
\usepackage{placeins}
\usepackage{tabularx}

\begin{document}
\title{A Computational Framework for the Acoustic Characterization of \textit{Suikinkutsu} (Water Harp Cave)}
\author{Haimes, Paul}
   {Ritsumeikan University}
   {haimes@fc.ritsumei.ac.jp}
\author{Uhde, Alarith}
   {Ritsumeikan University}
   {uhde@fc.ritsumei.ac.jp}
\author{Marutschke, Daniel Moritz}
   {Ritsumeikan University}
   {moritz@fc.ritsumei.ac.jp}
\begin{abstract} 
\textit{Suikinkutsu} are partially concealed acoustic devices traditionally installed adjacent to stone water basins (\textit{tsukubai}) in Japanese temple gardens. While previous research has examined their physical acoustics and cultural significance, computational methods for documenting and comparing their acoustic characteristics remain underexplored. This exploratory study introduces a computational framework for analyzing \textit{suikinkutsu} using spectral and temporal audio features extracted from field recordings. Rather than estimating the physical geometry of individual \textit{suikinkutsu}, the framework characterizes their acoustic profiles using a multidimensional set of computational descriptors, enabling systematic comparisons between recordings across different sites and installations. Audio recordings from nine \textit{suikinkutsu} located at temples, shrines, and gardens in the Kansai region of Japan were analyzed using a feature-extraction pipeline that incorporates Fast Fourier Transform (FFT), Mel-Frequency Cepstral Coefficients (MFCCs), Root Mean Square (RMS) energy, spectral centroid, spectral bandwidth, and spectral contrast. The extracted features indicate that, although the \textit{suikinkutsu} share broadly similar spectral and timbral characteristics, each exhibits a measurably distinct acoustic profile; a supplementary analysis of within-site and between-site variability suggests these profiles reflect a bounded range of characteristic resonant behaviour rather than a fixed acoustic signature. The proposed framework provides one of the first computational approaches for systematically characterizing and comparing \textit{suikinkutsu}, contributing a practical foundation for future research and the digital preservation of this distinctive form of Japanese environmental design.
\end{abstract}
\maketitle

\begin{center}
\setlength{\fboxrule}{1pt}
\fcolorbox{blue}{white}{\parbox{0.92\linewidth}{\color{blue}\footnotesize
\textbf{Author copy.} Please cite as: Haimes, P., Uhde, A., \& Marutschke, D. M. (2026; in press). A computational framework for the acoustic characterization of \textit{Suikinkutsu} (water harp cave). \textit{International Journal of Asia Digital Art and Design}.
}}
\end{center}

\section{Introduction}
\label{sec:intro}
This paper presents a computational framework for analyzing \textit{suikinkutsu} -- a term commonly translated as ``water harp cave'' -- a partially hidden feature found adjacent to stone water basins (\textit{tsukubai}) in several temple gardens throughout Japan. The device is, more rarely, also referred to as \textit{mizugoto}, meaning ``water harp''.
\begin{figure}[thb]
\centering
\includegraphics[width=0.9\linewidth]{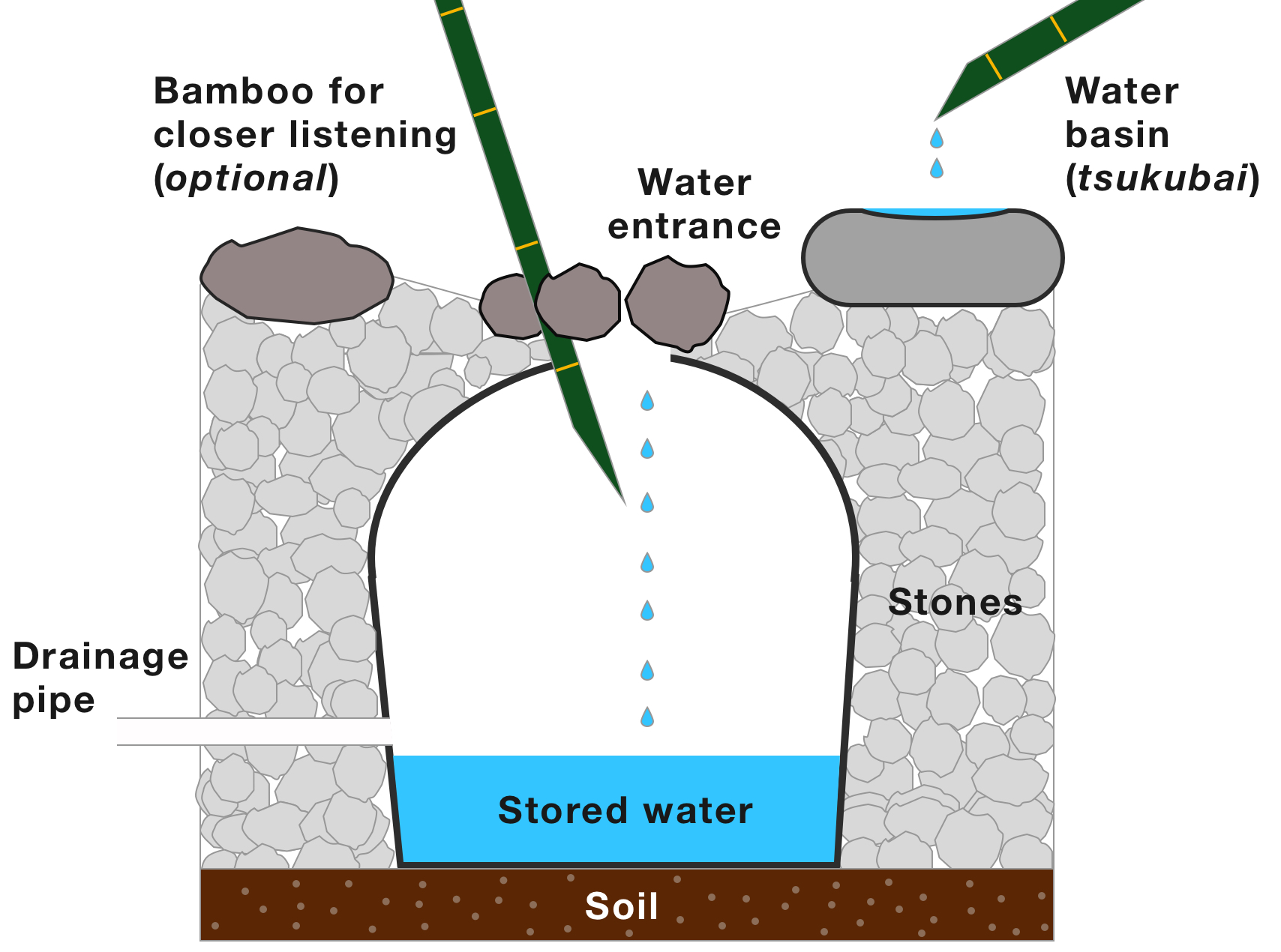}
\caption{Cross-section diagram of a \textit{suikinkutsu} \cite{Haimes2026Suikinkutsu}, based on the descriptions in \cite{Hirayama1958Suikinkutsu}.
Generally, \textit{suikinkutsu} are submerged, but some vessels exist that are above ground (e.g., Figure \ref{fig:photo}, right). }
\label{fig:diagram}
\end{figure}
As described in previous research, a \textit{suikinkutsu} consists of ``a buried ceramic or metal pot with a small opening at the top... Water drips into the pot, producing soft, resonant echoes, creating a soundscape that blurs the lines between the \textit{biophonic} (biological sounds) and \textit{anthrophonic} (human-made sounds)'' \cite{Haimes2026Suikinkutsu}. A piece of bamboo often protrudes from these devices, inviting visitors to listen carefully (Figures \ref{fig:diagram} and \ref{fig:photo}). While it is estimated that Japan has around 77,000 temples \cite{nationalgeographic2025temples}, the national \textit{suikinkutsu} database website (\textit{Zenkoku Suikinkutsu D\={e}tab\={e}su})\footnote{http://www.suikinkutsu.com}, though seemingly no longer updated, lists just 487 \textit{suikinkutsu} sites throughout the country.\\

\begin{figure*}[thb]
\centering
\includegraphics[width=1\linewidth]{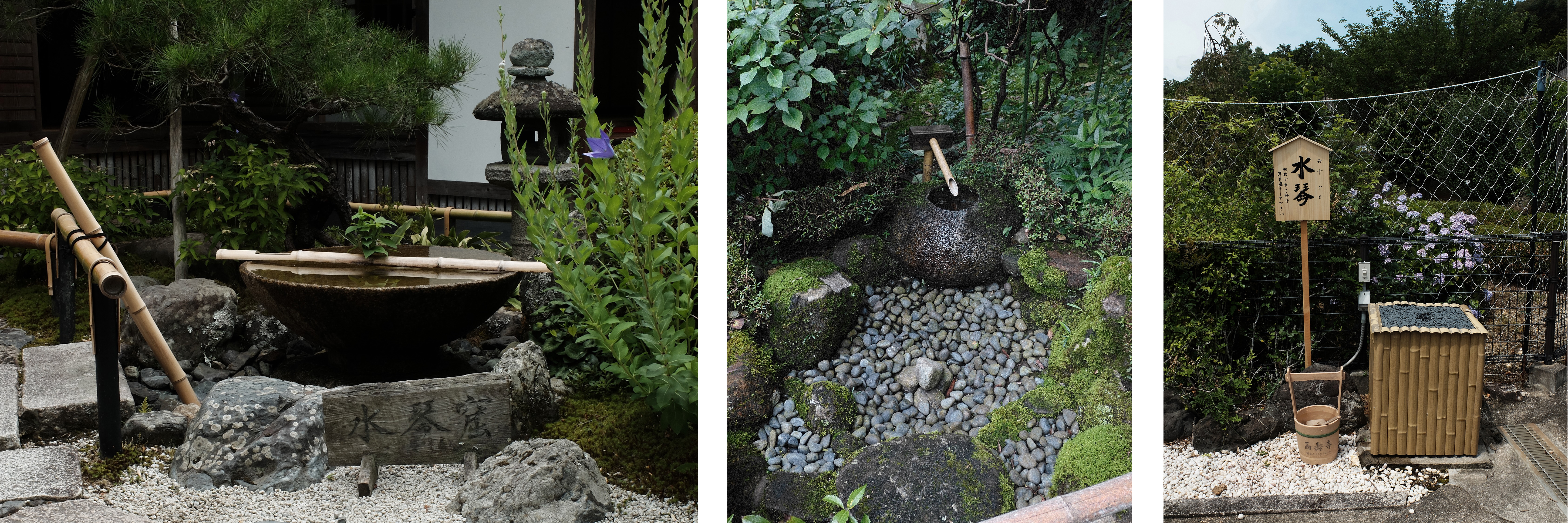}
\caption{Three contrasting \textit{suikinkutsu} configurations. Left: Enk\={o}ji, showing the common configuration with a bamboo pole for listening. Centre: One of two sites at Taima-dera, showing the stone basin and a small pile of rocks covering the water's entrance. Right: Saij\={u}ji, showing the above-ground vessel and wooden ladle (\textit{hishaku}) used to manually pour water, in place of a steady, elevated drip. Photographs by the author(s).}
 \label{fig:photo}
\end{figure*}
\subsection{Literature Review}
While the exact circumstances of their invention remain unclear, \textit{suikinkutsu} were constructed as a type of drainage device for excess water from stone water basins, and are said to have been in use as a garden feature since the Edo period (1603--1868) \cite{Hirayama1958Suikinkutsu}. Despite this history, relatively little research has examined \textit{suikinkutsu} in detail. By the time of Hirayama's 1958 survey, the device had fallen into such decline that it stood ``on the verge of near-total destruction, such that even its structure was about to be forgotten'' \cite{Hirayama1958Suikinkutsu}. It was mainly through the efforts of Hirayama -- who ``documented surviving examples and provided details of their structure'' \cite{Haimes2026Suikinkutsu} -- that they gained renewed attention during the latter half of the twentieth century. Nevertheless, they remain an underinvestigated phenomenon within Japanese aesthetics, and, in the authors' experience conducting fieldwork for this study, their scarcity relative to the number of temples in Japan may partly explain why relatively few Japanese people are even aware of their existence.\\
Early acoustic investigations by Kishizuka employed frequency spectrum (FFT) analysis to compare original and reconstructed \textit{suikinkutsu}, demonstrating that the characteristic sound is strongly influenced by the resonance of the vessel \cite{Kishizuka1991Suikinkutsu}. Similarly, Iwase compared the FFT spectra of \textit{suikinkutsu} with other traditional Japanese sound-producing devices -- including the \textit{koto}, \textit{shishi-odoshi} (deer scarer), and \textit{uguisubari} (nightingale floor) -- and, through simulation, modelled the \textit{suikinkutsu}'s distinctive tone as arising from a broadband water-drop excitation governed by the resonant modes of the vessel's internal cavity, also noting a comparatively long reverberation time of approximately 1.5--2 seconds in the dominant 1--5~kHz range \cite{Iwase2000Suikinkutsu}. \={O}izumi et al. later provided a foundational account of the physical acoustics of the \textit{suikinkutsu} \cite{Ooizumi2000Suikinkutsu}. Their work demonstrated that the characteristic sound arises from two related phenomena:
\begin{enumerate}
    \item Bubble vibration at the moment of impact of a water drop, which determines the frequency.
    \item Amplification by the natural resonance of the inverted vessel's air cavity.
\end{enumerate}
Building upon these findings, Watanabe investigated the combined effects of bubble excitation, vessel vibration and cavity resonance through theoretical and experimental modelling, further clarifying the acoustic mechanisms responsible for the distinctive sound of the \textit{suikinkutsu} \cite{Watanabe2004Suikinkutsu}. More recently, Fujita et al. used numerical simulation of the air--–water interaction to model sound radiation from the water surface itself, providing support for the resonance mechanisms identified in the preceding studies \cite{fujita2012simulation}. Similarly, Oku et al. used a computational fluid dynamics simulation to reproduce the sound pressure waveform of a single bubble, offering further support for the bubble resonance mechanism described above \cite{oku2018visualization}.\\
Recent work has also examined the cultural and aesthetic significance of the \textit{suikinkutsu} within Japanese gardens, exploring its role in environmental sound appreciation and drawing parallels between the subtle presence of the \textit{suikinkutsu} and the subtle, sustained listening required to appreciate Japanese ambient music (\textit{kanky\={o} ongaku}) \cite{Haimes2026Suikinkutsu}.  Complementing this, Cerwén examined \textit{suikinkutsu} as one of several sound-producing garden features within a broader soundscape design framework for Japanese gardens \cite{cerwen2019listening}. Cerwén's study noted that most of the Japanese gardens that contained soundscape elements were located in urban areas. This reality meant that ``noise from surrounding activities may be disturbing, but escaping from such noise into a quiet garden can make that garden seem all the more tranquil'' \cite{cerwen2019listening}.
\textit{Suikinkutsu}, along with other phenomena found in Japanese gardens, align with a broader tendency in Japanese aesthetics toward understated, easily overlooked forms of beauty \cite{Haimes2020Minimalism}.

\begin{table*}[t]
\centering
\caption{Recording setup and conditions for each of the nine \textit{suikinkutsu} sites. ``Takes'' is the total number of takes recorded at each site before discarding those interrupted mid-recording; three complete takes were retained per site for analysis (Section~\ref{sec:datacollection}).}
\label{tab:sites}
\scriptsize
\setlength{\tabcolsep}{3pt}
\begin{tabularx}{\textwidth}{
>{\raggedright\arraybackslash}p{2.2cm}
>{\raggedright\arraybackslash}p{0.9cm}
>{\raggedright\arraybackslash}p{1.8cm}
>{\raggedright\arraybackslash}p{0.9cm}
>{\raggedright\arraybackslash}p{0.6cm}
>{\raggedright\arraybackslash}p{1.1cm}
>{\raggedright\arraybackslash}p{0.7cm}
>{\raggedright\arraybackslash}p{0.5cm}
>{\raggedright\arraybackslash}p{1.4cm}
>{\raggedright\arraybackslash}X}
\hline
\textbf{Location} &
\textbf{Site Type} &
\textbf{Date / Time} &
\textbf{Device} &
\textbf{Bamboo} &
\textbf{Mic Placement} &
\textbf{Pouring} &
\textbf{Takes} &
\textbf{Weather} &
\textbf{Background Noise} \\
\hline
Enk\={o}ji temple, Kyoto & Temple & 23 Jun 2026, 13:42--14:24 & TP-7 & Yes & Mouth of bamboo & Auto & 5 & Partly sunny & Birds/insects, passers-by \\
Eikand\={o} (Zenrinji) temple, Kyoto & Temple & 30 Jun 2026, 14:30--14:51 & TP-7 & No & Water entrance & Auto & 6 & Sunny & Passers-by, traffic \\
Taiz\={o}in, a sub-temple of My\={o}shinji, Kyoto & Temple & 24 Mar 2026, 14:10--14:36 & iPhone SE & Yes & Mouth of bamboo & Auto & 3 & Overcast & Birds/insects, passers-by, nearby stream \\
Saij\={u}ji temple, Kyoto & Temple & 3 Jul 2026, 10:20--10:56 & TP-7 & No & Water entrance & Manual & 4 & Sunny & Manual pouring, passers-by \\
Unry\={u}in temple, Kyoto & Temple & 3 Jul 2026, 12:27--13:05 & TP-7 & Yes & Mouth of bamboo & Auto & 4 & Very sunny & Passers-by, traffic \\
Kyoto Botanical Gardens & Public garden & 4 Jul 2026, 09:46--10:32 & TP-7 & Yes & Water entrance & Auto & 5 & Partly sunny & Birds/insects, passers-by, traffic \\
Miwa shrine, Nara & Shrine & 7 Jul 2026, 13:50--14:22 & TP-7 & Yes & Mouth of bamboo & Auto & 9 & Heavily overcast, humid & Birds/insects, passers-by, rituals \\
Taima-dera (1), Nara & Temple & 7 Jul 2026, 11:42--12:05 & TP-7 & No & Water entrance & Auto & 4 & Overcast & Birds/insects, traffic \\
Taima-dera (2), Nara & Temple & 7 Jul 2026, 11:42--12:05 & TP-7 & No & Water entrance & Auto & 4 & Overcast & Birds/insects, traffic \\
\hline
\end{tabularx}
\end{table*}
\subsection{Motivation for the Study}
This exploratory study extends previous research by introducing a computational framework for documenting and comparing the acoustic characteristics of \textit{suikinkutsu} using audio analysis. Instead of estimating the physical geometry of individual \textit{suikinkutsu}, this research examines which acoustic features remain relatively consistent across recorded audio signals and can thus be used to characterize specific examples. Given that all recordings were captured \textit{in situ} under variable and only partially controlled conditions, this study treats the resulting acoustic profiles as exploratory descriptions of recording-level characteristics rather than as claims about fixed, intrinsic, and fully reproducible properties of each \textit{suikinkutsu}; Section~\ref{sec:variability} reports an analysis intended to assess the extent to which these profiles nonetheless remain stable across repeated recordings at the same site.\\
This study's objective is specifically to characterize the \textit{suikinkutsu} as a sound-producing structure -- that is, to describe acoustic properties attributable to the vessel's resonance and construction -- rather than to document the broader \textit{in situ} soundscape experience of the garden settings in which \textit{suikinkutsu} are found, a topic addressed more directly by prior soundscape-oriented research \cite{cerwen2019listening}. Ambient environmental sound is nonetheless present in all field recordings. Because no objective method was available in the field to measure or separate ambient noise from vessel resonance, this paper does not treat environmental sound as an object of study in its own right, or claim to isolate it with certainty; instead, likely instances of environmental noise are flagged and discussed through informed interpretation where evident in the recordings (Sections~\ref{sec:acoustic-char}, \ref{subsec:sites}, and \ref{sec:resonant}), rather than through a rigorously validated threshold (Section~\ref{sec:limitations}).\\
Estimating the geometry of \textit{suikinkutsu} is challenging because reliable information about their dimensions is rarely available. Temple records are often unavailable or inaccessible, and knowledgeable staff are difficult to locate. Additionally, \textit{suikinkutsu} are relatively uncommon and are not present at all temples in Japan, which limits the number of sites available for study. As a result, this research is situated at the intersection of \textit{acoustic analysis} and \textit{digital heritage}, illustrating how computational audio analysis can document and compare \textit{suikinkutsu} when physical measurements are inaccessible. In this context, \textit{acoustic characterization} denotes the representation of individual \textit{suikinkutsu} using a multidimensional set of computational audio descriptors that enable comparison between recordings.\\
This framing situates the present study within a broader recognition that the acoustic properties of a heritage site or object constitute a legitimate, and often overlooked, dimension of cultural heritage in their own right, distinct from -- though complementary to -- the physical structure that produces them \cite{Brezina2013Acoustics}. Where digital heritage practice has more commonly relied on visual or architectural documentation, such as photogrammetry or 3D scanning, computational audio descriptors of the kind employed here offer a comparable systematic approach for documenting sound-producing heritage objects, aligning with recent proposals to use structured sound classification as a practical tool for heritage management \cite{ParkerSpennemann2022Classifying}, and with recent experimental and numerical work characterizing the acoustics of individual heritage structures as a form of intangible cultural heritage in its own right \cite{Ricciutelli2024SanBiagio}. This approach is particularly well suited to \textit{suikinkutsu}: Japan has an existing precedent of formally recognizing everyday and environmental sounds as heritage worth conserving, exemplified by the Ministry of the Environment's \textit{100 Soundscapes of Japan} initiative \cite{MOE1996Soundscapes}, which itself includes two named \textit{suikinkutsu} -- one in Gunma Prefecture and one in Gifu Prefecture -- among its selections, out of the several hundred known \textit{suikinkutsu} sites nationwide (Section~\ref{sec:intro}). The computational framework proposed here offers one practical means of extending this kind of documentation to the many \textit{suikinkutsu} that have not individually received such recognition.
\section{Method}
\subsection{Data Collection}
\label{sec:datacollection}
Field recordings were made using either a \textit{Teenage Engineering TP-7}\footnote{https://teenage.engineering/products/tp-7} portable recorder or an \textit{iPhone SE} (2nd gen) across nine temple gardens, shrine gardens, and other gardens in the Kansai region of central Japan, all open to the public, summarized in Table~\ref{tab:sites}.

The device microphones were placed close to the mouth of the bamboo, with the exception of Eikand\={o}, Saij\={u}ji, Taima-dera (1), and Taima-dera (2), which lacked bamboo poles, and Kyoto Botanical Gardens, where a bamboo pole was present but the microphone was instead positioned close to the exposed water entrance. At Saij\={u}ji, water was poured manually, rather than via a steady, elevated drip as at the other sites (see Figure~\ref{fig:photo}). At Miwa, the \textit{suikinkutsu} was noticeably quiet at the source despite the presence of a dedicated bamboo pole, and, as a busy, frequently visited shrine, background noise was more difficult to avoid than at the other sites.\\
At each site, three usable takes, each approximately 30 seconds in duration, were recorded within a single visit lasting approximately 32 minutes, on days with generally clear or overcast weather to avoid interference from rain (see Table~\ref{tab:sites} for the conditions noted at each site), yielding approximately 13.5 minutes of raw field recording across the nine sites in total. At a few sites, additional takes were recorded because an in-progress take was interrupted (e.g.\ by a passing visitor) and subsequently discarded, so that three complete, uninterrupted takes could be retained per site. Microphone placement was kept as consistent as possible across takes at a given site; because sites were sometimes visited by other members of the public concurrently, however, exact placement could not always be replicated between takes. As described in Section~\ref{sec:preprocessing}, each 30-second take was subsequently trimmed to a 10-second segment for analysis, yielding 30 seconds of analyzed audio per site (270 seconds in total across all twenty-seven takes). For the site-level comparisons reported in Section~\ref{subsec:results}, the clearest of the three retained takes at each site -- judged by the lowest level of extraneous ambient noise -- was manually selected as that site's representative recording.
\subsection{Audio Pre-Processing}
\label{sec:preprocessing}
Each audio file was manually processed as follows in \textit{RX Audio Editor}\footnote{https://www.izotope.com/products/rx-advancediZotope} software:
\begin{itemize}
\item Each 30-second take was trimmed to a 10-second segment; the segment containing the least background noise was manually selected.
\item Normalized to target peak levels of 0.0 dBFS to ensure consistent volume across all audio files.
\item Exported as a 16-bit 44.1 kHz mono WAV file.
\item Files were checked to confirm they did not contain any clipping (distortion caused by signal levels exceeding the recording system's maximum amplitude).
\end{itemize}
\renewcommand{\arraystretch}{1.5}
\begin{table*}[t]
\centering
\caption{Relationship between the computational descriptors, their acoustic interpretation, and prior work. Each of these is a common audio analysis feature available in \textit{Librosa}, though \textit{SciPy} was also used to filter and estimate the frequency spectrum for FFT visualization (Figure \ref{fig:fft}).}
\label{tab:features}
\small
\begin{tabularx}{\textwidth}{
>{\raggedright\arraybackslash}p{3.2cm}
>{\raggedright\arraybackslash}X
>{\raggedright\arraybackslash}X
>{\raggedright\arraybackslash}p{3.2cm}}
\hline
\textbf{Descriptor} &
\textbf{Acoustic Measure} &
\textbf{Physical Interpretation} &
\textbf{Related Work} \\
\hline
Fast Fourier Transform (FFT)
&
Dominant resonant frequencies
&
Primary cavity resonance
&
Kishizuka; Iwase; \={O}izumi et al.; Fujita et al. \\
Root Mean Square (RMS)
&
Resonance decay
&
Resonance persistence and damping behaviour
&
Iwase; \={O}izumi et al. (implicit) \\
Spectral Bandwidth
&
Distribution of resonant energy
&
Spread of resonant frequencies
&
Watanabe \\
Spectral Contrast
&
Prominence of resonant peaks
&
Prominence of individual resonant frequencies
&
Kishizuka \\
Spectral Centroid
&
Overall spectral energy distribution
&
Brightness of the resonant spectrum
&
Watanabe \\
Mel-Frequency Cepstral Coefficients (MFCC)
&
Compact generalized acoustic profile approximating human hearing
&
Overall timbral signature
&
General-purpose descriptor (implemented via \textit{Librosa}) \\
\hline
\end{tabularx}
\end{table*}
\subsection{Feature Extraction}
Acoustic features were extracted from the recordings using the \textit{Librosa} \cite{mcfee2015} Python library -- a widely-used framework for extracting spectral and temporal descriptors commonly used in computational audio analysis:
\begin{equation*}
\mathbf{f}_i = \Phi(x_i(t))
\label{eq:feature_vector}
\end{equation*}
where $x_i(t)$ is the recorded audio signal from the $i$-th \textit{suikinkutsu}, $\Phi$ denotes the feature extraction process, and $\mathbf{f}_i$ is the resulting acoustic feature vector whose components are summarized in Table~\ref{tab:features}, which also shows the relationship between the selected descriptors, how they can be physically interpreted, and relevant prior studies.
\begin{figure}[thb]
\centering
\includegraphics[width=1\linewidth]{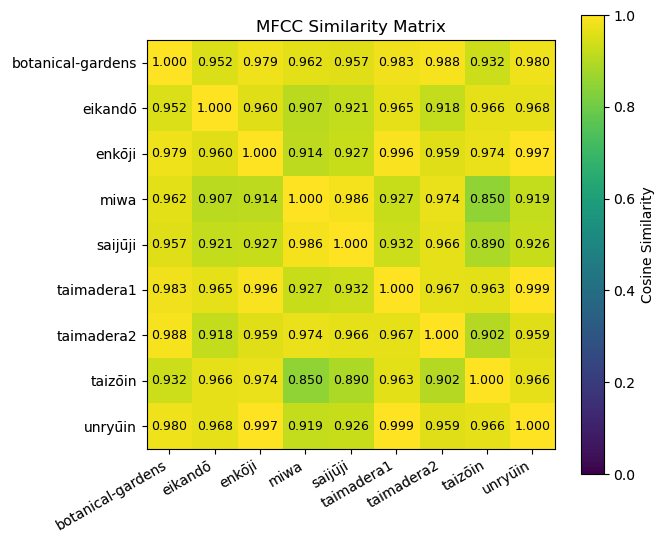}\caption{Pairwise acoustic similarity between the recorded \textit{suikinkutsu}, computed using cosine similarity of the mean MFCC feature vectors. Higher values indicate recordings with more similar overall timbre.}
\label{fig:similarity}
\end{figure}
Computational audio descriptors were selected to correspond to the principal acoustic mechanisms and characteristics investigated in previous studies, including cavity resonance, spectral characteristics, and resonance decay. Where possible, descriptors were chosen to reflect those used or implied by earlier work, with additional general-purpose spectral descriptors included to provide a more comprehensive characterization. None of these individual descriptors is novel in isolation: each is a standard, general-purpose descriptor already available in \textit{Librosa}, and prior studies have already applied computational spectral analysis, including FFT, to \textit{suikinkutsu} at individual sites \cite{Kishizuka1991Suikinkutsu, Iwase2000Suikinkutsu, Ooizumi2000Suikinkutsu, fujita2012simulation}. The framework proposed here is accordingly not offered as a new signal-processing or music-information-retrieval technique, but as a literature-grounded combination of these descriptors (Table~\ref{tab:features}), applied consistently and comparatively across field recordings collected from multiple sites for the purpose of digital heritage documentation -- a combination and application that, to our knowledge, has not previously been reported for \textit{suikinkutsu}. While FFT analysis captures dominant resonant frequencies of each \textit{suikinkutsu}, the combination of complementary spectral and temporal descriptors provides a more comprehensive characterization of their acoustic signatures. For visual clarity, the frequency spectra shown in Figure~\ref{fig:fft} were additionally processed, using the \textit{SciPy} Python library \cite{Virtanen2020SciPy}, with a high-pass filter (30~Hz cutoff) to remove sub-audible rumble and DC offset, and estimated using Welch's method rather than a single raw FFT, producing a smoother spectral estimate than the unfiltered computation implied elsewhere in this section. Figures~\ref{fig:similarity}--\ref{fig:fft} were generated using the Python library \textit{Matplotlib} \cite{Hunter2007}.
\subsection{Within-Site and Between-Site Variability}
\label{sec:variability}
To assess whether the acoustic differences reported between sites (Section~\ref{subsec:results}) reflect stable characteristics of individual \textit{suikinkutsu} rather than variation attributable to recording conditions or the manual selection of a representative take (Section~\ref{sec:datacollection}), pairwise cosine similarity was additionally computed across all three retained takes per site ($N = 27$ recordings), rather than the nine manually-selected representative recordings used elsewhere in this paper. Each pairwise comparison was classified as \textit{within-site} (both takes from the same site; $n = 27$ pairs) or \textit{between-site} (takes from different sites; $n = 324$ pairs). Because pairs sharing a recording are not independent of one another, statistical significance was assessed using a permutation test: site labels were randomly reassigned across the 27 recordings 10{,}000 times, and the observed difference between mean within-site and mean between-site similarity was compared against the resulting null distribution.
\section{Results}
\subsection{Timbral Similarity}
\label{sec:timbral}
Despite being recorded at different sites, with different vessel construction, materials, and dimensions, the \textit{suikinkutsu} examined here share broadly similar spectral and timbral characteristics. Figure~\ref{fig:similarity} presents the cosine similarity matrix derived from the mean MFCC vectors of each recording. Higher similarity values indicate recordings with more closely related spectral envelopes, while lower values suggest greater differences in their overall timbral characteristics.
\begin{figure}[thb]
\centering
\includegraphics[width=1\linewidth]{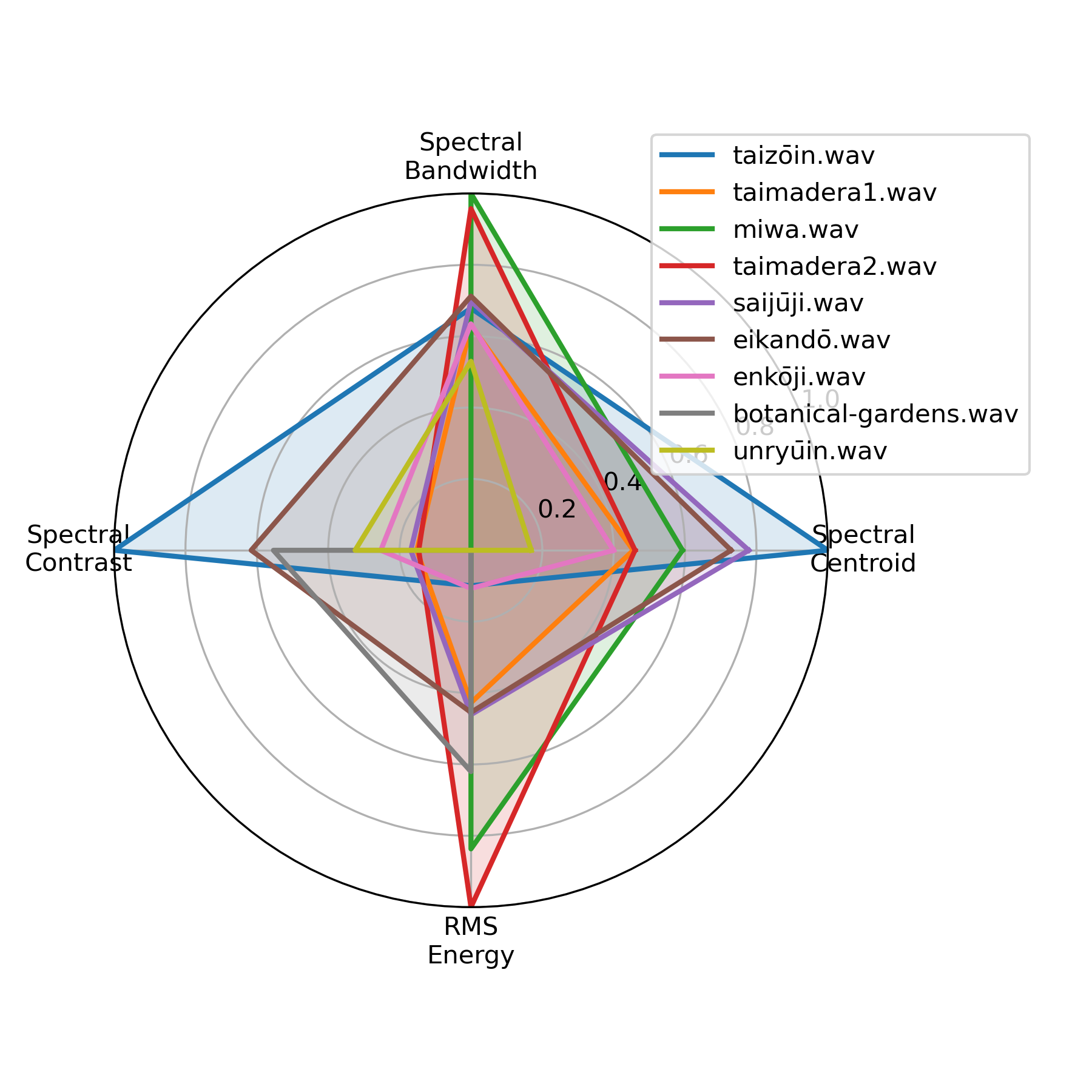}\caption{Normalized acoustic profiles derived from the extracted computational descriptors. Each polygon represents the multidimensional feature vector of an individual \textit{suikinkutsu}, illustrating differences in spectral centroid, bandwidth, contrast, and RMS energy.}
\label{fig:radar}
\end{figure}
\begin{figure}[thb]
\centering
\includegraphics[width=1\linewidth]{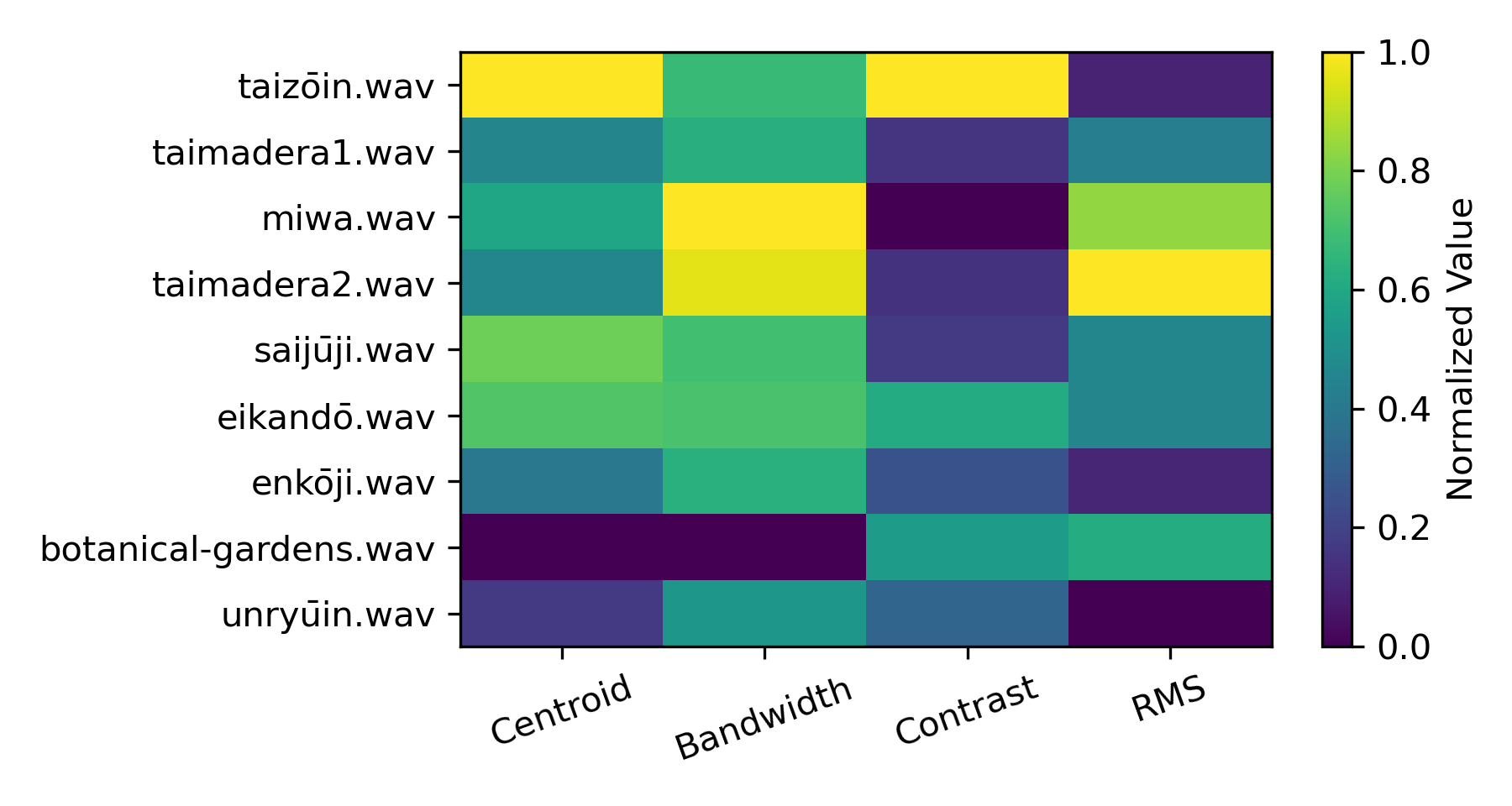}\caption{Comparison of the normalized computational descriptors extracted from each \textit{suikinkutsu}. The feature matrix provides a compact visual summary of the similarities and differences between the acoustic signatures of the nine recordings.}
\label{fig:heatmap}
\end{figure}
The matrix (Figure~\ref{fig:similarity}) reveals that all nine recordings share broadly similar overall timbral characteristics, with pairwise cosine similarity values falling between 0.850 and 0.999. Notably, the lowest pairwise similarity in the entire matrix occurs between Miwa and Taiz\={o}in (0.850), which nonetheless remains a comparatively high similarity score in absolute terms. This suggests that even the most acoustically dissimilar \textit{suikinkutsu} in the sample exhibit a high degree of overall timbral similarity despite site-specific differences. Although the proposed descriptors reveal measurable differences between recordings (Section~\ref{subsec:results}), their overall spectral envelopes remain highly similar.
\begin{figure}[thb]
\centering
\includegraphics[width=1\linewidth]{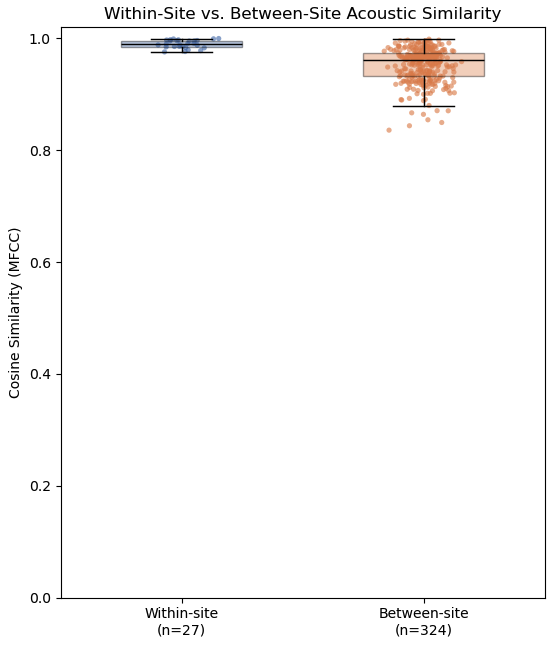}
\caption{Distribution of pairwise MFCC cosine similarity for within-site ($n=27$; three takes recorded during a single visit to each of the nine sites) and between-site ($n=324$) recording pairs. Within-site pairs show consistently higher and more tightly clustered similarity than between-site pairs.}
\label{fig:variability}
\end{figure}
As a further check on whether these site-level differences might instead reflect the manual selection of a representative recording per site (Section~\ref{sec:datacollection}), pairwise similarity was also computed across all three unselected takes per site (Figure~\ref{fig:variability}). Within-site pairs ($M = 0.99$) were consistently more similar than between-site pairs ($M = 0.95$), a separation unlikely to have arisen by chance (permutation test, $p < .0001$; Section~\ref{sec:variability}). The magnitude of this gap is modest relative to the overall timbral similarity noted above, consistent with individual \textit{suikinkutsu} exhibiting a bounded range of characteristic resonant behaviour rather than a fixed acoustic fingerprint -- a distinction returned to in Section~\ref{sec:conclusion}.
\subsection{Acoustic Characterization}\label{subsec:results}
\label{sec:acoustic-char}
Despite this overall timbral similarity, descriptor-level differences become apparent when the recordings are compared using the full multidimensional feature vectors. Figure~\ref{fig:radar} presents the resulting acoustic profiles, while Figure~\ref{fig:heatmap} provides a feature comparison matrix. Whereas Figure~\ref{fig:radar} emphasizes the multidimensional acoustic profile of each recording, Figure~\ref{fig:heatmap} facilitates descriptor-by-descriptor comparison across all nine sites.
\begin{figure*}[t!]
\centering
\includegraphics[width=1.005\linewidth]{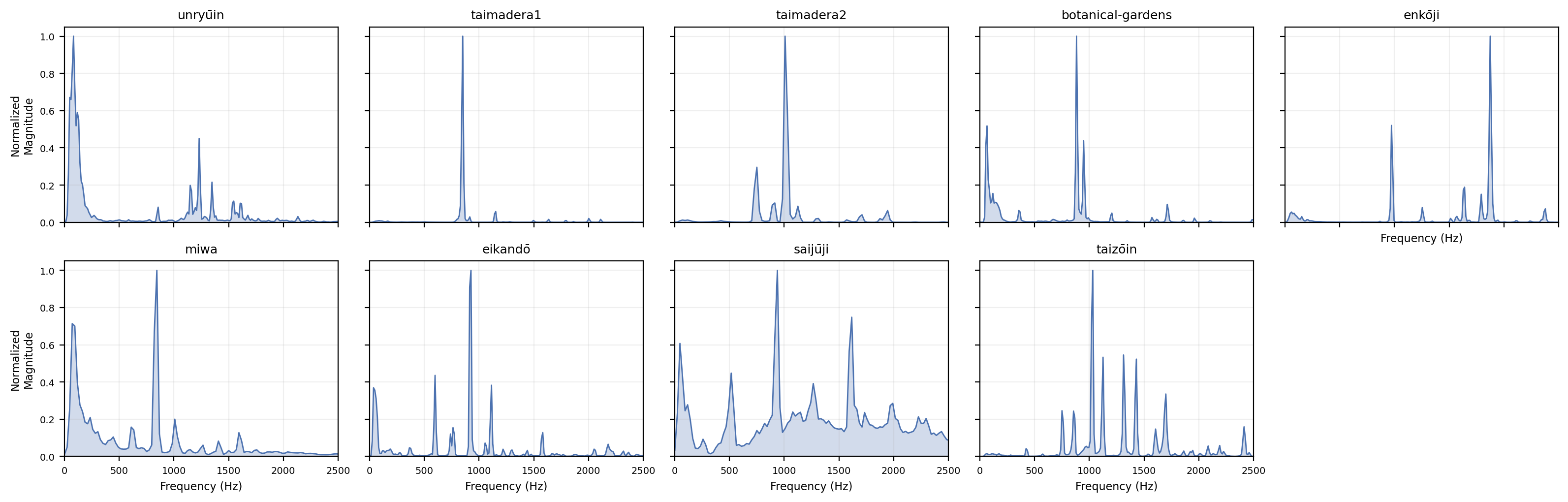}
\caption{Normalized FFT spectra of the nine \textit{suikinkutsu} recordings, shown as small multiples and ordered by spectral centroid (lowest to highest). The dominant resonant peaks correspond to the cavity resonance described in the related literature, while variations between recordings suggest differences in the acoustic characteristics of individual \textit{suikinkutsu}. Some low-frequency activity visible in a subset of recordings is attributable to ambient environmental noise rather than vessel resonance, as discussed in the text.}
\label{fig:fft}
\end{figure*}
The selected descriptors reveal measurable descriptor-level differences among the recorded \textit{suikinkutsu}. \\Taiz\={o}in exhibits the highest spectral centroid among the recordings, together with a bandwidth still more moderate than Miwa's or Taimadera2's, suggesting a brighter, more open spectral response with greater emphasis on higher frequencies. Unry\={u}in, by contrast, exhibits the lowest spectral centroid among the recordings (Figure~\ref{fig:fft}), consistent with a darker, lower-frequency-weighted resonance. Eikand\={o} and Enk\={o}ji occupy intermediate positions within the descriptor space, exhibiting characteristics between these two extremes.
\subsection{Site-Specific Acoustic Divergences}\label{subsec:sites}
Although the recordings are broadly similar overall, several sites exhibit somewhat low MFCC similarity scores that merit closer examination. \\One such example is Saij\={u}ji, which shares relatively low timbral similarity with Taiz\={o}in (0.890). Saij\={u}ji differs from most other sites in two important respects: it lacks the bamboo pole used to channel sound toward the microphone, and water is introduced by manual pouring rather than a steady, elevated drip. These differences likely influence the timing, amplitude, and spectral envelope of the recorded water impacts, contributing to its distinctive descriptor profile. The absence of a bamboo pole alone is unlikely to explain this divergence, however, as Eikand\={o} -- also lacking bamboo -- occupies a more central position within the descriptor space. These observations therefore suggest that the manually poured water source is likely a contributing factor. This interpretation is further supported by Saij\={u}ji's relatively elevated spectral activity in the 300--600~Hz range (Figure~\ref{fig:fft}), consistent with the broader-band, less periodic impacts expected from manual pouring. In the absence of verified physical measurements, however, these explanations remain interpretive.\\
The Miwa recording forms the least similar pair in the dataset with Taiz\={o}in (0.850), and is also among Miwa's lower pairings with Eikand\={o} (0.907). Unlike the example discussed above, this recording was made using the standard bamboo-mouth configuration, making recording geometry an unlikely explanation for its distinctiveness. Instead, the source itself was noticeably quiet, while the shrine's high visitor traffic made background noise difficult to avoid. Because all recordings were normalized to a common peak level, the quieter source retained proportionally more ambient noise after normalization than recordings made under quieter conditions. This offers a plausible explanation for Miwa's comparatively distinct acoustic profile more broadly, not only its divergence from Taiz\={o}in, although confirmation would require formal noise-floor analysis or verified physical measurements.\\
The Kyoto Botanical Gardens recording also exhibits comparatively low similarity with several other sites, including Taiz\={o}in (0.932). Unlike Saij\={u}ji, the site includes a bamboo pole, but the recording was made adjacent to the water entrance rather than at the bamboo opening. Despite this microphone placement, it remains highly similar to conventionally recorded sites such as Enk\={o}ji (0.979) and Unry\={u}in (0.980). It also shares a distinctly high similarity with Saij\={u}ji (0.957), despite the latter lacking a bamboo pole and relying on manual water pouring, suggesting that recording position and water-delivery method do not necessarily drive large timbral divergence on their own. The recording also contains pronounced energy below approximately 100~Hz (Figure~\ref{fig:fft}), corresponding to traffic noise from a nearby road that was directly observed during field recording and likely contributes to this site's markedly low similarity with Taiz\={o}in, independently of the \textit{suikinkutsu}. Nevertheless, the low spectral centroid and bandwidth values for the Kyoto Botanical Gardens in Figure~\ref{fig:heatmap} suggest that at least some of this divergence from several sites is due to the characteristics of the \textit{suikinkutsu} itself.\\
Collectively, these observations suggest that the proposed descriptor set captures genuine acoustic differences between individual \textit{suikinkutsu} while reflecting the actual recording conditions. Rather than representing a limitation alone, this sensitivity reflects the realities of documenting heritage soundscapes \textit{in situ}, where environmental factors inevitably contribute to the recorded acoustic signature.
\subsection{Resonant Characteristics}
\label{sec:resonant}
These descriptor-level differences are also reflected at the level of the raw frequency spectra. Figure~\ref{fig:fft} compares the frequency spectra of the nine recordings. Although all recordings exhibit pronounced resonant peaks characteristic of the \textit{suikinkutsu} (most recordings exhibit their strongest resonant activity between approximately 600 and 2000 Hz, with the exception of Unry\={u}in, Miwa, Eikand\={o}, and Kyoto Botanical Gardens, which had a comparatively high amount of activity under 400 Hz), differences are evident in both the location and prominence of the dominant resonances. Several of these peaks cluster near 1000~Hz, within the 1--3~kHz range in which Iwase similarly reported pronounced spectral peaks \cite{Iwase2000Suikinkutsu}. Rather than reflecting vessel resonance, low-frequency activity is more likely attributable to ambient background noise -- including traffic, nearby conversation, and wind -- directly noted during recording at these four sites (see also the discussion of traffic noise at Kyoto Botanical Gardens in Section~\ref{subsec:sites}). Bearing this in mind, it is clear that \textit{suikinkutsu} are generally active within a broadly consistent frequency range across sites, with the low-frequency activity observed at these four recordings being most plausibly attributable to environmental factors rather than genuine divergence in resonant behaviour. These observations are consistent with previous studies identifying cavity resonance as the principal mechanism governing sound production. RMS energy measurements indicate variation in resonance persistence between sites, suggesting differences in damping behaviour (i.e., how quickly the resonance decays over time; see the RMS axis in Figure~\ref{fig:radar}). Longer resonance decay times may reflect differences in cavity geometry, material properties, or other acoustic factors, although the absence of physical measurements prevents attribution to any single cause. It is worth noting that the boundary between genuine low-frequency resonance and ambient noise cannot be established with certainty. Therefore, the frequency ranges discussed above should be read as indicative rather than definitive.
Taken together, these results suggest that \textit{suikinkutsu} exhibit broadly similar acoustic characteristics across different sites regardless of differences in vessel dimensions, materials, and construction, while still exhibiting measurable, recording-level differences that the proposed framework is able to capture. Whether these differences correspond to stable, reproducible properties of each \textit{suikinkutsu} or are partly attributable to recording conditions is considered further in Section~\ref{sec:variability} and Section~\ref{sec:conclusion}.
\section{Conclusion, Limitations, and Future Work}
\label{sec:conclusion}
This exploratory study introduced a computational framework for the acoustic characterization of \textit{suikinkutsu} through audio feature extraction. Building on previous research into the physical acoustics of \textit{suikinkutsu}, a set of descriptors was extracted from recordings collected at nine sites in the Kansai region of Japan. Rather than possessing a fixed or clearly delineated acoustic signature -- an expectation difficult to sustain given that all recordings were captured \textit{in situ} under variable environmental conditions -- individual \textit{suikinkutsu} appear to exhibit a bounded range of characteristic resonant behaviour, arising from fixed physical constraints such as vessel geometry, cavity volume, and construction materials, even as moment-to-moment excitation (e.g., drip timing, manual pouring, ambient noise) remains inherently variable (Section~\ref{subsec:results}). The proposed framework provides a practical method for documenting and comparing these culturally significant sound-producing structures, supporting their preservation as part of Japan's intangible sonic heritage.
\subsection{Limitations}
\label{sec:limitations}
Consistent with the exploratory framing adopted in this study (Section~\ref{sec:intro}), the primary limitation is the limited number of recordings and the lack of verified physical measurements for the individual \textit{suikinkutsu}. As a result, the research emphasizes acoustic characterization rather than precise geometric estimation. Furthermore, as all recordings were captured on-site rather than under controlled acoustic conditions, several were subject to uncontrollable environmental variables -- including ambient traffic, conversation, and wind -- which likely contributed to some of the low-frequency divergence observed between sites (Section~\ref{sec:acoustic-char}). Peak normalization (Section~\ref{sec:preprocessing}), while necessary to ensure a consistent overall signal level across recordings for the purpose of comparison, also proportionally raises the relative contribution of background noise in comparatively quiet source recordings, as observed for Miwa (Section~\ref{subsec:sites}); this interaction was not independently controlled for or corrected. Relatedly, although the within-site and between-site variability analysis (Section~\ref{sec:variability}) indicates that acoustic differences between sites exceed the variation observed across takes recorded during a single visit, this analysis has two further limits worth stating explicitly. First, it cannot rule out the possibility that some between-site differences are also attributable to each site having been visited on a different day, and therefore under somewhat different ambient conditions; disentangling site-level acoustic identity from such day-to-day environmental variation would require repeated visits to each site across a range of conditions, which we identify as a direction for future work. Second, because the three takes retained at a given site were recorded with the same device and water-delivery method, and the same or similar microphone placement, low within-site variance is consistent both with a stable acoustic identity for the vessel and with the same recording equipment and procedure simply having been used repeatedly at that site; the present design does not fully separate these possibilities, and doing so would require varying recording conditions independently of site. Separately, all comparisons in this study are among \textit{suikinkutsu} recordings; no non-\textit{suikinkutsu} recordings were collected for comparison. We do not treat this as a limitation of the argument advanced here: the claims made throughout this paper are comparative and within-category, characterizing and comparing \textit{suikinkutsu} recordings to one another for the purpose of digital heritage documentation, and do not assert that \textit{suikinkutsu} are always acoustically distinguishable from all other resonant-vessel or dripping-water sources in general; establishing that broader, between-category claim is not something this analysis needs to support. Given the nature of the sites at which \textit{suikinkutsu} are installed --- fixed, embedded garden features rather than portable objects --- recording them under a controlled acoustic setting is inherently difficult, and this study was accordingly conducted \textit{in situ}. In any case, a dedicated control comparison is of limited practical value here: an acoustically unrelated submission to the crowd-sourced platform proposed as future work would be expected to score poorly in the initial MFCC analysis and reveal itself automatically (Section~\ref{sec:futurework}). Varying recording conditions independently of site, discussed above, remains the genuine direction for future work arising from this analysis. Notably, this also meant that a precise boundary between genuine low-frequency vessel resonance and ambient environmental noise could not be established; the attributions offered in Section~\ref{sec:acoustic-char} therefore reflect informed interpretation rather than a rigorously validated threshold. Future work may benefit from supplementary noise-floor measurements or recording protocols better able to isolate the \textit{suikinkutsu} from surrounding environmental sound. However, it is worth bearing in mind that ambient background noise is an inseparable part of the \textit{suikinkutsu} experience.
\subsection{Future Work}
\label{sec:futurework}
Onset-based descriptors for quantifying drip timing were also explored but were not incorporated into the present framework, as the distinction between individual impacts and sustained resonance proved unreliable, particularly across recordings exhibiting both discrete dripping and continuous water flow. Future work may investigate more robust onset-detection methods validated against manually annotated recordings.\\
Future work could also investigate the relationship between acoustic signatures and physical geometry, should architectural measurements become available. A further direction for future work is the development of a dedicated, modern platform for crowd-sourced documentation of \textit{suikinkutsu}. This could lead to larger datasets and the incorporation of additional descriptors. Data-driven clustering methods may facilitate the automatic classification of \textit{suikinkutsu} based on acoustic similarity. As noted in Section~\ref{sec:intro}, the existing national database appears not to have been updated in over two decades; a bilingual website and mobile application could allow visitors to submit photographs and audio recordings of \textit{suikinkutsu} they encounter, along with location and accessibility information, helping to verify which sites remain active. Submitted recordings would need to be manually reviewed to ensure sufficient audio quality before inclusion in any dataset. In receiving publicly sourced recordings, which are likely to vary considerably in quality and recording conditions, a minimum acceptable cosine similarity to existing verified recordings -- derived from the pairwise comparisons reported in this study (Section~\ref{sec:acoustic-char}) -- could additionally serve as a useful automated screening threshold, helping to ensure that submitted audio remains recognisably of a \textit{suikinkutsu} prior to inclusion in any dataset. As stated previously, beyond manual review, an acoustically unrelated recording submitted to such a platform would quickly reveal itself through a low MFCC similarity score below this threshold (see also Section~\ref{sec:limitations}). Such a platform could substantially expand the scale of data available for analysis, enabling more robust machine learning approaches than the comparative framework presented here, which was necessarily constrained by the small number of recordings collected through manual fieldwork. This would, however, constitute a separate undertaking beyond the scope of the present study.

\section{Acknowledgements}\textit{Claude Code} was used to assist in coding the visualizations used in Figures 3--7. Thank you to the anonymous reviewers for their comments on two earlier versions of this paper.
Special thanks also to Mrs Machiko Haimes who assisted with recording at some of the sites used in this research.

\end{document}